# Passenger Route and Departure Time Guidance under Disruptions in Oversaturated Urban Rail Transit Networks

**Siyu Zhuo**
School of Traffic and Transportation
Key Laboratory of Transport Industry of Comprehensive Transportation Theory, Ministry of Transport
Beijing Jiaotong University, Beijing, 100044, China
Email: syzhuo@bjtu.edu.cn

**Xiaoning Zhu**
School of Traffic and Transportation
Key Laboratory of Transport Industry of Comprehensive Transportation Theory, Ministry of Transport
Beijing Jiaotong University, Beijing, 100044, China
Email: xnzhu@bjtu.edu.cn

**Pan Shang, Corresponding Author**
School of Traffic and Transportation
Key Laboratory of Transport Industry of Comprehensive Transportation Theory, Ministry of Transport
Beijing Jiaotong University, Beijing, 100044, China
Email: shangpan@bjtu.edu.cn

**Zhengke Liu**
School of Transportation Science and Engineering
Beijing Key Laboratory for Cooperative Vehicle Infrastructure System and Safety Control
Beihang University, Beijing, 100191, China
Email: zkliu@buaa.edu.cn

*Siyu Zhuo, Xiaoning Zhu, Pan Shang and Zhengke Liu*

**ABSTRACT**

The urban rail transit (URT) system attracts many commuters with its punctuality and convenience. However, it is vulnerable to disruptions caused by factors like extreme weather and temporary equipment failures, which greatly impact passengers' journeys and diminish the system's service quality. In this study, we propose targeted travel guidance for passengers at different space-time locations by devising passenger rescheduling strategies during disruptions. This guidance not only offers insights into route changes but also provides practical recommendations for delaying departure times when required. We present a novel three-feature four-group passenger classification principle, integrating temporal, spatial, and spatio-temporal features to classify passengers in disrupted URT networks. This approach results in the creation of four distinct solution spaces based on passenger groups. A mixed integer programming model is built based on individual level considering the First-in-First-out (FIFO) rule in oversaturated networks. Additionally, we present a two-stage solution approach for handling the complex issues in large-scale networks. Experimental results from both small-scale artificial networks and the real-world Beijing URT network validate the efficacy of our proposed passenger rescheduling strategies in mitigating disruptions. Specifically, when compared to scenarios with no travel guidance during disruptions, our strategies achieve a substantial reduction in total passenger travel time by 29.7% and 50.9% respectively, underscoring the effectiveness in managing unexpected disruptions.

**Keywords:** Urban rail transit network, Disruption, Passenger classification, Route guidance, Departure time guidance

## INTRODUCTION

### Background

Urban rail transit (URT) systems play a pivotal role in providing efficient and sustainable transportation services, especially amid rapid urbanization and population growth (*1*). Despite their significance, these essential systems are susceptible to disruptions caused by adverse weather, aging infrastructure, ongoing expansion, and near-capacity operations. These disruptions, varying from minor delays to complete shutdowns, significantly impact passenger travel, resulting in delays, overcrowding, trip cancellations, and substantial economic losses (*2*). Effectively addressing the adverse effects of these disruptions on passengers necessitates the implementation of strategies to guide them through the aftermath of such disruptions.

Recent efforts have been devoted to comprehending and managing the impact of disruptions on URT networks, focusing on implementing control measures from two primary perspectives. On the supply side, URT operators aim to optimize operational adjustments by modifying the timetables of relevant lines (*3*). Meanwhile, on the demand side, attention has been given to facilitating rescheduling passengers' travel, involving route choice and departure time adjustments. However, it is essential to acknowledge that existing research has predominantly emphasized rescheduling trains from the operator's viewpoint, which might not always align with ensuring passengers' convenience and satisfaction. Moreover, operational adjustments, such as temporary timetable modifications, may face challenges in timeliness during unexpected disruptions, making demand-side approaches potentially more effective in such scenarios (*4*).

Demand-side approaches, such as offering passengers personalized route choices and advising them on optimal departure times, offer significant benefits for individual passengers. These approaches help minimize the effects of disruptions, improve travel efficiency, and enhance the overall passenger experience. However, it's essential to recognize that guidance can vary significantly depending on passengers' diverse space-time locations. For instance, passengers outside stations during disruptions might prefer alternative transportation modes, whereas those already within stations may choose to wait for the next available train, expecting a return to normal service. Therefore, accurately categorizing passengers according to their locations and devising tailored travel rescheduling strategies are of utmost importance (*5*).

Furthermore, disruptions at a certain segment frequently trigger ripple effects that disseminate to other stations and lines of the URT systems, leading to a deterioration of the overall service quality (*2*). However, tackling URT disruptions in a large-scale network is an exceptionally challenging task due to the complex structure of the URT system and the diverse choices exhibited by passengers. Existing strategies proposed for managing URT disruptions are predominantly localized, primarily focusing on partial adjustments for specific lines or stations (*6*), which may be of poor quality on the network level. Therefore, there is a need for comprehensive optimization of the overall network's service level at a macro level.

Given these challenges, this study delves into network-wide passenger guidance during severe disruptions, such as temporary blockages in a segment of a URT line. By classifying passengers according to their spatial and temporal positions at the onset of disruption, our goal is to craft tailored travel guidance. This guidance not only suggests alternative routes but also advises on optimal departure times. We develop personalized rescheduling strategies at the individual level and propose a two-stage solution approach to address the large-scale disruptions in URT networks. These travel rescheduling strategies empower URT operators to effectively manage disruptions by providing passengers with timely guidance, thereby bolstering the resilience and operational reliability of URT systems.

### Literature review

Disruption management in railway and URT systems has been a topic of extensive research. Various studies have explored the management of disruptions in conventional railway lines, Zhan et al. (*4*) proposed a rolling horizon approach to address partial segment blockages in high-speed railway lines. Zhu et al. (*7*) constructed a timetable rescheduling model where flexible stopping and short-turning are innovatively integrated with three other dispatching measures. In the context of URT systems, scholars also make several explorations to alleviate the impact of disruptions. Lusby et al. (*8*) presented a branch-and-price algorithm to solve rolling stock rescheduling for suburban lines. Wang et al. *(6)* studied the integrated train rescheduling and rolling stock circulation planning for complete blockage situations in URT lines, focusing on timetable deviations, cancellations, and headway deviations.

However, these studies mostly pay attention to the approaches from supply-side optimization, including the adjustment of the line planning, the train timetabling, and the rolling stock scheduling, which have overlooked demand-side strategies, such as the guidance of passenger behaviors, which is also crucial in solving real-world disruptions. Recognizing this importance, some researchers have incorporated passengers' guidance in their disruption management strategies. Like supply-side-oriented studies, the optimization objects can also lie in two sides (i.e., railway and URT systems) when to guide demand-side passengers. As summarized as **Table 1** below, railway systems and URT systems exhibit distinct characteristics in terms of infrastructure, operational dynamics, and passenger information systems. Railway systems typically feature centralized stations and fixed routes, often connecting major cities or regions with longer distances covered. This necessitates the use of pre-planned schedules and timetables, with passenger information systems and electronic message boards at stations providing updates on accidents, train schedules, and weather conditions *(9)*. On the other hand, URT systems are characterized by distributed stations in urban areas, shorter distances between stations, and more frequent services with flexible routing options. Passengers in URT systems rely heavily on smartphone navigation apps for real-time updates on train schedules, route options, and traffic conditions *(10)*. These systems also employ congestion management strategies, such as dynamic routing and scheduling, along with passenger information displays at stations, to provide timely information and enhance passenger experience *(11)*. Hence, it suggests that guiding passengers to change routes can effectively mitigate the impact of disruptions in URT systems. For example, Mo et al. *(12)* introduced a probabilistic framework that leverages smart card data to infer passenger response behavior during unplanned rail service disruptions. Results indicate that during scenarios with high redundancy, most affected passengers (69.51%) chose to make route changes to mitigate delays. Wang et al. *(13)* systematically investigated the optimization of bus bridging service design and passenger assignment during disruptions in urban rail transit. The results demonstrate that the bus bridging routes generated can significantly reduce both operator-oriented and passenger-oriented costs.

**TABLE 1 Characteristic comparison between railway and urban rail transit systems**

| Characteristic | Railway system | Urban rail transit system |
|---|---|---|
| Passenger information systems | Electronic message board | Navigational apps on smartphones |
| Ticket price | Expensive | Cheap |
| Distance between stations | Long | Short |
| Correspondence between tickets and seats | One-to-one, reservation required | One-to-many, no reservation required |
| Recovery time | Long | Short |
| Route substitutability | Low | High |
| Transfer time | Long | Short |

While the studies mentioned above providing personalized and optimal route choices during unexpected disruptions to guide passengers' route changes, selecting appropriate departure times can also be attributed to avoiding congestion and delays caused by disruptions in URT systems. Li et al. *(14)* proposed a mixed logit model to describe the departure time choice of metro passengers, considering the endogeneity of price. They found that travelers prefer to depart earlier and pay a lower fare if they need to adjust their departure time. However, the study of adjusting passenger departure times when disruptions occur has not yet received adequate attention and thorough examination.

Moreover, because passengers' choices during disruptions are greatly influenced by their spatial and temporal locations, scholars have attempted to study passenger choices in more detail by categorizing passengers when disruptions occur. For instance, Sun et al. *(2)* utilized tap-in and tap-out data to estimate the impacts of disruptions on passenger travel time and delays in URT networks. They classified passengers into three categories: 1) "missed" passengers who left the system, 2) passengers who took detours, and 3) passengers who experienced delays but continued their journeys. Similarly, Mo et al. (*12*) categorized passenger choices under disruptions into 15 classes based on their locations at the time of disruption, analyzing behavioral characteristics for each category using a probabilistic choice model.

Additionally, unexpected disruptions in URT systems exhibit a high degree of the butterfly effect, spreading rapidly throughout the network. However, due to the enormous number of OD (origin-destination) pairs and the presence of multiple transfer stations, devising effective passenger guidance strategies becomes extremely challenging among the network, especially for a large one. Consequently, existing research has primarily focused on optimizing single line or station. For example, Bešinović et al. (*3*) introduced a novel integrated disruption management model aimed at simultaneously rescheduling trains and controlling passenger flows during disruptions of a line within the Beijing metro system. Besides, Yin et al. (*15*) introduced a two-stage stochastic optimization model aimed at rescheduling the timetable and serving passengers delayed by disruptions. Real-world case studies, based on historical data from a Beijing metro line, validate the effectiveness of the proposed approach in reducing passengers' travel time. Hence, further exploration into network-level optimization, including many lines and stations, remains to be addressed.

**Contribution**

In the face of sudden disruptions within URT networks, passengers may face delays. Providing personalized travel guidance for passengers in different space-time locations, including route and departure time adjustments, is crucial to mitigating these disruptions and improving the overall passenger experience. This study aims to provide such guidance, contributing to improving the overall passenger experience during disruptions. The contributions can be summarized as follows.

1) Four decision groups are proposed to formulate customized travel rescheduling strategies for passengers using the three-feature four-group classification principle, which incorporates temporal, spatial, and spatiotemporal features.

2) An individual-level mixed integer model is developed to design customized rescheduling strategies for passengers belonging to four groups. These strategies encompass not only route guidance but also departure time guidance, an aspect that has been relatively underexplored in current research. Considering the oversaturated situations, the FIFO principle is also depicted.

3) A two-stage solution approach encompassing the passenger classification and passenger updating stages is developed to address the computational challenges of large-scale problems.

4) Experimental tests are conducted on several networks of different scales, which illustrate the effectiveness of the proposed travel rescheduling strategy for passengers in addressing unexpected disruptions. URT operators can enhance their capacity to handle disruptions efficiently by categorizing passengers based on their space-time trajectories when unexpected disruptions occur.

**MODEL FORMULATION**

**Three-feature Four-group Passenger Classification Principle**

While individual passengers tend to be concerned only about their travel changes, URT operators are often more focused on efficiently guiding and dispersing different types of passengers throughout the entire network. To accurately understand the flow of passengers in the URT system after an unexpected disruption and to provide targeted travel rescheduling guidance for different passenger categories, we have developed a three-feature four-group classification principle. This approach helps us summarize the possible passenger choices that may occur when disruptions happen in the URT network, as depicted in **Figure 1**.

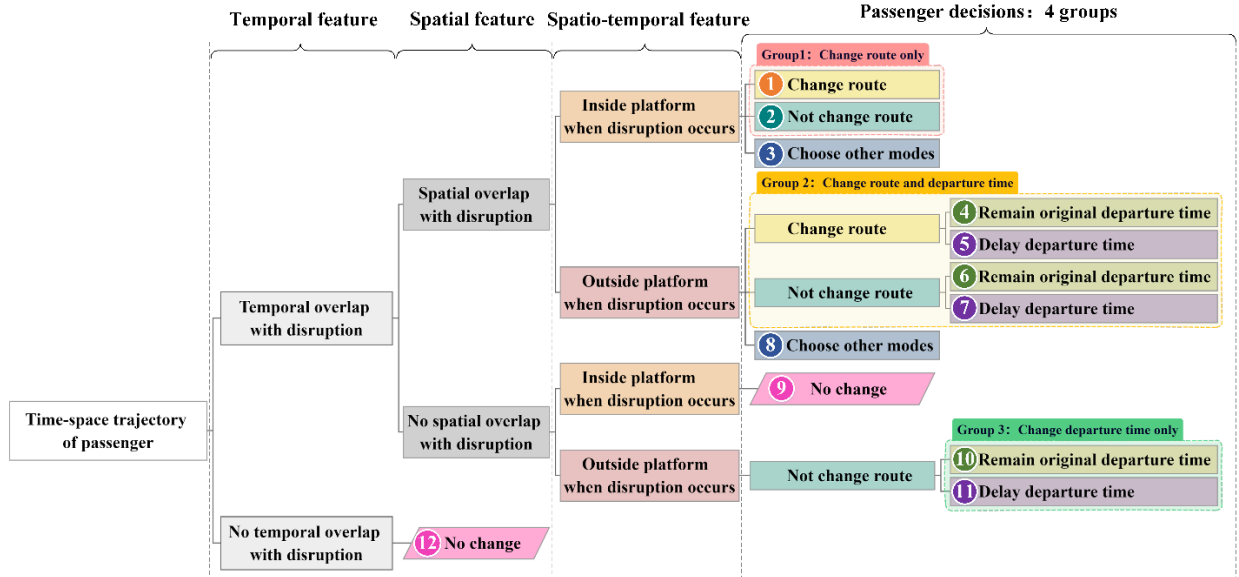

**Figure 1 Three-feature four-group classification principle of passenger possible choices**

The three-feature four-group classification principle identifies 12 potential passenger choices. The first feature categorizes passengers into "temporal overlap with disruption" and "no temporal overlap with disruption" based on whether their travel time aligns with the disruption interval. Passengers outside this interval belong to the latter group.

Passengers are further classified by the spatial feature into "spatial overlap with disruption" and "no spatial overlap with disruption" depending on whether their routes intersect the affected segment. Those passing through the affected segment are in the former group, while others are in the latter.

The third feature considers the spatio-temporal aspect. Passengers' choices vary based on their locations when the disruption occurs. Those outside the station (e.g, at home) can opt for alternative transportation or delay their departure, while those inside must either wait or change routes. This classification feature first divides potentially affected passengers into those inside the platform and those outside the URT system at the time of the disruption.

During a disruption in the URT system, certain stations or segments may become inaccessible, hindering train movement. Passengers inside the stations with travel routes affected by the disruption have three options: 1) change routes using alternative lines within the URT system; 2) maintain their current routes and wait at their present locations until the disrupted section reopens; or 3) exit the URT system and switch to other transportation modes.

For passengers outside the URT system during the disruption, an additional option is available, specifically delaying their departure time, especially for passengers at home. Therefore, passengers outside the URT system have five options: 4) change routes and keep their original departure times; 5) change routes and delay their departure times; 6) maintain their current routes and original departure times; 7) maintain their current routes and delay their departure times; or 8) switch directly to other transportation modes.

Similarly, passengers not directly affected by the disruption (those who have "no spatial overlap with disruption") may make different choices based on their locations at the time of the disruption. To minimize disruptions, passengers whose routes do not intersect with the disrupted section will keep their original routes. Passengers inside the stations will not change routes or departure times, leading to option 9) no change. Passengers outside the stations will also maintain their original routes but can delay departure to avoid congestion, resulting in options 10) maintain routes with original departure times and 11) maintain routes with delayed departure times. Lastly, passengers unaffected by the timing of the disruption will also choose option 12) no change. To facilitate clarity, each passenger choice is assigned a specific ID for reference in **Figure 2**.

In total, there are twelve categories of passenger choices. By classifying these twelve categories based on the decisions passengers can make, i.e., whether they can change their routes and/or delay their departure times after the disruptions, all choices can be summarized into four groups of decisions:

**Group 1:** Change route only
**Group 2:** Change route and departure time
**Group 3:** Change departure time only
**Group 4:** No change

The four-group classification principle will guide the establishment of path pools and passenger departure time windows in the subsequent model development. Its advantages will be demonstrated through the passenger classification algorithm, enhancing the efficiency of passenger categorization.

Notably, the validity of this classification principle has been proved through several studies by other scholars. Mo et al. (*12*) for instance, have introduced a comparable classification principle and furnished empirical substantiation based on survey data to corroborate the practical utility of this methodology. This classification principle has been empirically demonstrated to possess a broad scope of applicability across various transit systems, and it can be readily customized and employed in alternative public transport networks, providing valuable insights into potential passenger travel choices. Currently, there is existing literature that has expanded upon this foundation to estimate the impact of passengers under unforeseen disrupted conditions in public transit systems, such as Chen et al. (*16*). Sun et al. (*2*) also employed data from automated fare collection (AFC) facilities, along with data from historical network disruptions, to categorize abnormal passengers into three characteristic types and further dividing them into four distinct groups.

Furthermore, a standard within the transportation industry of China called 'Specifications for Developing a Response Plan to Urban Rail Transit Operation Emergencies' explicitly states that when managing unforeseen disruptions in URT systems, emergency plans should be tailored based on the disruption's location and its scope of impact, taking into account the passengers' locations. As a result, our classification principle, which considers the relationship between the location and duration of unforeseen disruptions and passengers' spatiotemporal trajectories, aligns with the industry standards and is supported by relevant empirical research (*2*, *12*, *16*). For those interested in more details, further information can be found in Mo et al. (*12*) for a comprehensive exploration.

## Passenger Route and Departure Time Guidance Model

*Notations*
For readers' convenience, the sets, indices, parameters, and decision variables throughout this study are summarized in **Table 2**.

**TABLE 2 Notations**

| Notations | Definitions |
|---|---|
| **Sets** | |
| $N$ | Set of nodes in an URT network |
| $L$ | Set of links in an URT network |
| $L^p$ | Set of links in route $p$ in an URT network |
| $L^a$ | Set of unfinished links for passenger group $a$ |
| $L^{fifo}$ | Set of entry and transfer arcs in an URT network, where passengers need to comply with the FIFO principle while waiting in queues |
| $q_{ijtt'}$ | Set of waiting passengers in order on space–time entry/transfer arc $(i, j, t, t')$ |
| $o_{ijtt'}$ | Set of in-vehicle passengers on space–time running arc $(i, j, t, t')$ |
| $V$ | Set of vertices in a space-time network |
| $E$ | Set of arcs in a space-time network |
| $A^{g1}$ | Set of passengers classified into Group 1 |

| $A^{g2}$ | Set of passengers classified into Group 2 |
|---|---|
| $A^{g3}$ | Set of passengers classified into Group 3 |
| $A^{g4}$ | Set of passengers classified into Group 4 |
| $A$ | Set of passengers, $A = A^{g1} \cup A^{g2} \cup A^{g3} \cup A^{g4}$ |
| $T$ | Set of time intervals |
| $W$ | Set of origin-destination (OD) pairs |
| $P^{w(a)}$ | Set of feasible routes for OD pair $w$ of passenger $a$ |
| $\Delta(a)$ | Set of feasible departure time points of passenger $a$ |
| **Indices** | |
| $i, j$ | Index of nodes/stations, $i, j \in N$ |
| $(i, j)$ | Index of links between two adjacent nodes, $(i, j) \in L$ |
| $(i, t)$ | Index of space-time vertices, $(i, t) \in V$ |
| $(i, j, t, t')$ | Index of space-time edges/arcs, $(i, j, t, t') \in E$ |
| $a$ | Index of passengers, $a \in A$ |
| $t, t'$ | Index of time intervals, $t, t' \in T$ |
| $o_a$ | Index of the origin node of passenger $a$ (the station node with disruption), $o_a \in N$ |
| $d_a$ | Index of the destination node of passenger $a$, $d_a \in N$ |
| $o_a^{other}$ | Index of the origin node for other transportation mode of passenger $a$, $o_a^{other} \in N$ |
| $w(a)$ | Index of the OD pair of a passenger $a$, $w(a) \in W$ |
| $p$ | Index of routes, $p \in P^{w(a)}$ |
| $p^{ini}$ | Index of the initial path of passenger, $p^{ini} \in P^{w(a)}$ |
| $r$ | Index of the dummy origin node |
| $s$ | Index of the dummy destination node |
| **Parameters** | |
| $TT_{ij}$ | Travel time for physical link $(i, j)$ |
| $\mathcal{E}$ | Total passenger travel time |
| $c_{ijtt'}$ | Travel time for space-time arc $(i, j, t, t')$ |
| $C_{pt}$ | Time-dependent route travel time in route $p$ with departure time $t$ |
| $\tau$ | Maximum time that departure time can be delayed of passenger $a$ |
| $M$ | Assumed large value as an auxiliary parameter |
| $cap_{ijtt'}$ | Capacity of the trains on arc $(i, j, t, t')$ |
| $\Delta t$ | Maximum time length that passenger can delay the departure |
| $\delta_{ij}^p$ | 0-1 parameter for the relationship between a route and link; it is equal to 1 if link $(i, j)$ belongs to route $p$; otherwise, it is 0 |
| $\pi_{board}$ | Number of boarding passengers |
| $len(q_{ijtt'})$ | Number of waiting passengers on space–time entry/transfer arc $(i, j, t, t')$ |
| $len(o_{ijtt'})$ | Number of in-vehicle passengers on space–time running arc $(i, j, t, t')$ |
| $\theta_{ijtt'}$ | Binary indicator which equals 1 if space–time running arc $(i, j, t, t')$ is constructed based on the timetable; otherwise, it equals 0 |
| **Decision Variables** | |
| $x_{ijtt'}^a$ | **Passenger assignment variables** <br> Variable is 1 if passenger $a$ is assigned to arc $(i, j, t, t')$; otherwise, it is 0 |
| $z_{pt}^a$ | **Route and departure-time guidance variables** <br> Variable is 1 if passenger $a$ departs at time interval $t$ and is guided to choose route $p$; otherwise, it is 0 |

*Model motivation*

To provide targeted route and departure time guidance for passengers of various groups, we have established the following mixed integer programming model. To capture the oversaturated situations, especially the "failure to board" phenomena of passengers in real-life scenarios, we have imposed tight capacity constraint and stipulated a FIFO rule for passengers. The detailed objective function and constraints are as follows.

*Objective function*

$$\min Z = \sum_{i,j,t,t':(i,j,t,t')\in E} \sum_{a\in A} x^a_{ijtt'} c_{ijtt'} \tag{1}$$

*Constraints*

$$\sum_{t:t\in\Delta(a)} x^a_{ro(a)0t} = 1 \ \ \forall a \in A \tag{2}$$

$$\sum_{t:t\in T} x^a_{d(a)st|T|} = 1 \ \ \forall a \in A \tag{3}$$

$$\sum_{i,t:(i,j,t,t')\in E} x^a_{ijtt'} - \sum_{j',t'':(i,j',t',t'')\in E} x^a_{jj't't''} = 0 \ \ \forall a \in A, \forall (j,t') \in V/\{(r,0),(s,|T|)\} \tag{4}$$

$$\sum_{a\in A} x^a_{ijtt'} \le cap_{ijtt'} \ \ \forall (i,j,t,t') \in E \tag{5}$$

$$\sum_{t'',t''':t''<t} x^{a'}_{ijt''t'''} + \sum_{t'',t''':t''\ge t'} x^{a'}_{jj't''t'''} \le (1-x^a_{ijtt'})\times M + 1 \ \ \forall (i,j) \in L^{fifo}, \forall (i,j,t,t') \in E \tag{6}$$

$$\sum_{p\in P^{w(a)}} \sum_{t\in\Delta(a)} z^a_{pt} = 1 \ \ \forall a \in A \tag{7}$$

$$x^a_{ijtt'} \times \delta^p_{ij} \le z^a_{pt} \ \ \forall a \in A^{g1}, \forall p \in P^{w(a)}, \forall t \in \{t\}, \forall (i,j) \in L^p, \forall (i,j,t,t') \in E \tag{8}$$

$$x^a_{ijtt'} \times \delta^p_{ij} \le z^a_{pt} \ \ \forall a \in A^{g2}, \forall p \in P^{w(a)}, \forall t \in \{t|t \le t + \Delta t\}, \forall (i,j) \in L^p, \forall (i,j,t,t') \in E \tag{9}$$

$$x^a_{ijtt'} \times \delta^p_{ij} \le z^a_{pt} \ \ \forall a \in A^{g3}, \forall p \in \{p^{ini}\}, \forall t \in \{t|t \le t + \Delta t\}, \forall (i,j) \in L^p, \forall (i,j,t,t') \in E \tag{10}$$

$$x^a_{ijtt'} \times \delta^p_{ij} \le z^a_{pt} \ \ \forall a \in A^{g4}, \forall p \in \{p^{ini}\}, \forall t \in \{t\}, \forall (i,j) \in L^p, \forall (i,j,t,t') \in E \tag{11}$$

$$x^a_{ijtt'} = \{0,1\} \tag{12}$$

$$z^a_{pt} = \{0,1\} \tag{13}$$

The model aims to minimize passenger travel time as shown in **Equation 1**, which serves as the objective function. For each passenger $a$, the space-time flow balance constraints can be expressed as **Equations 2-4**, respectively. To effectively characterize the typical features of oversaturated passenger flows and queueing phenomena in URT systems, **Equation 5** establishes that the total number of passengers passing through any space-time arc $(i,j,t,t')$ must not exceed the capacity limit of this arc, which depends on the train capacity passing through this arc. Due to train capacity limitations, not all passengers waiting on the platform can board the trains, especially during peak periods with high demands, some passengers may be left behind by the first train and must wait for the subsequent train services. In real-world scenarios, overtaking behavior is generally not considered in traffic assignment models that involve time-dependent flows (*17, 18*). Therefore, the space-time passenger flow should adhere to the FIFO rule to prevent arbitrary, unrealistic, or physically impractical deviations (*19, 20*). At any time $t$, the sequence of passengers going through any section should adhere to FIFO rule, which means that a passenger who arrives at a queue early should leave the queue earlier instead of arriving late. The FIFO constraint can be expressed by **Equation 6**. This equation only holds for $x_{ijtt'}^a = 1$, and otherwise, the associated big-M value is turned on and thereby eliminates the constraints. **Equation 7** states that passenger $a$ should choose one departure time $t$ in the departure time set $\Delta(a)$ and one route from candidate route set $P^{w(a)}$. **Equation (8-11)** define how the choice of route and departure time varies for four different passenger groups and specify the solution space available for assigning passengers. Specifically, passenger $a$ can arrive at origin station $o(a)$ at time $t$ through path $p$ only if the departure time $t$ is chosen by passenger $a$ ($z_{pt}^a = 1$). In addition, the feasible path set $P^{w(a)}$ and the feasible departure time set $\Delta(a)$ exhibit variations based on different passenger groups. For passenger Group 1, which is limited to route modifications, the feasible path set, denoted as $P^{w(a)}$, encompasses all alternative routes between the passenger's origin and destination, represented as $P^{w(a)} = od : (o_a, d_a)$. The associated departure time set is $\{t\}$ for this group. In contrast, for passengers in Group 2, with the flexibility to adjust both travel routes and departure times, the feasible path set $P^{w(a)}$ maintains an identical one to that of Group 1. However, the departure time set $\Delta(a)$ extends from $\{t\}$ to $\{t | t \le t + \Delta t\}$. For Group 3, which concentrates solely on adjusting the departure time, the feasible path set includes only the passengers' initial path, denoted as $P^{w(a)} = \{p^{ini}\}$, with a departure time set $\Delta(a) = \{t | t \le t + \Delta t\}$. Lastly, Group 4, characterized by passengers who make no alterations to their travel routes or departure times, has a feasible path set $P^{w(a)} = \{p^{ini}\}$, while the $\Delta(a)$ remains $\{t\}$. **Equation 12** and **Equation 13** define the decision variables of the proposed model.

This proposed passenger route and departure time guidance model tailors travel rescheduling strategies for passengers of diverse groups following unforeseen disruptions by customizing feasible path sets and departure time windows specific to each group.

## TWO-STAGE SOLUTION APPROACH BASED ON ITERATION

### Overall Framework

Managing and guiding passengers' travel after disruptions on URT networks present significant challenges due to the complexity of network structures, diverse passenger travel choices, and unpredictable events. This is especially daunting in scenarios with high passenger volumes and extensive network coverage. Hence, this study proposes an iterative two-stage solution approach to address the problem of passenger guidance after disruptions based on the schedule-based passenger assignment in URT networks. The two stages and the specific algorithms involved in each stage are as follows:

- ✓ **Stage 1 → Classification：** Passenger Classification Algorithm (**A1**)
- ✓ **Stage 2 → Updating：** Passenger Updating Algorithm (**A2**)

The overall framework of the two-stage solution approach and the relationships between the inputs and outputs of each stage are depicted in **Figure 2**.

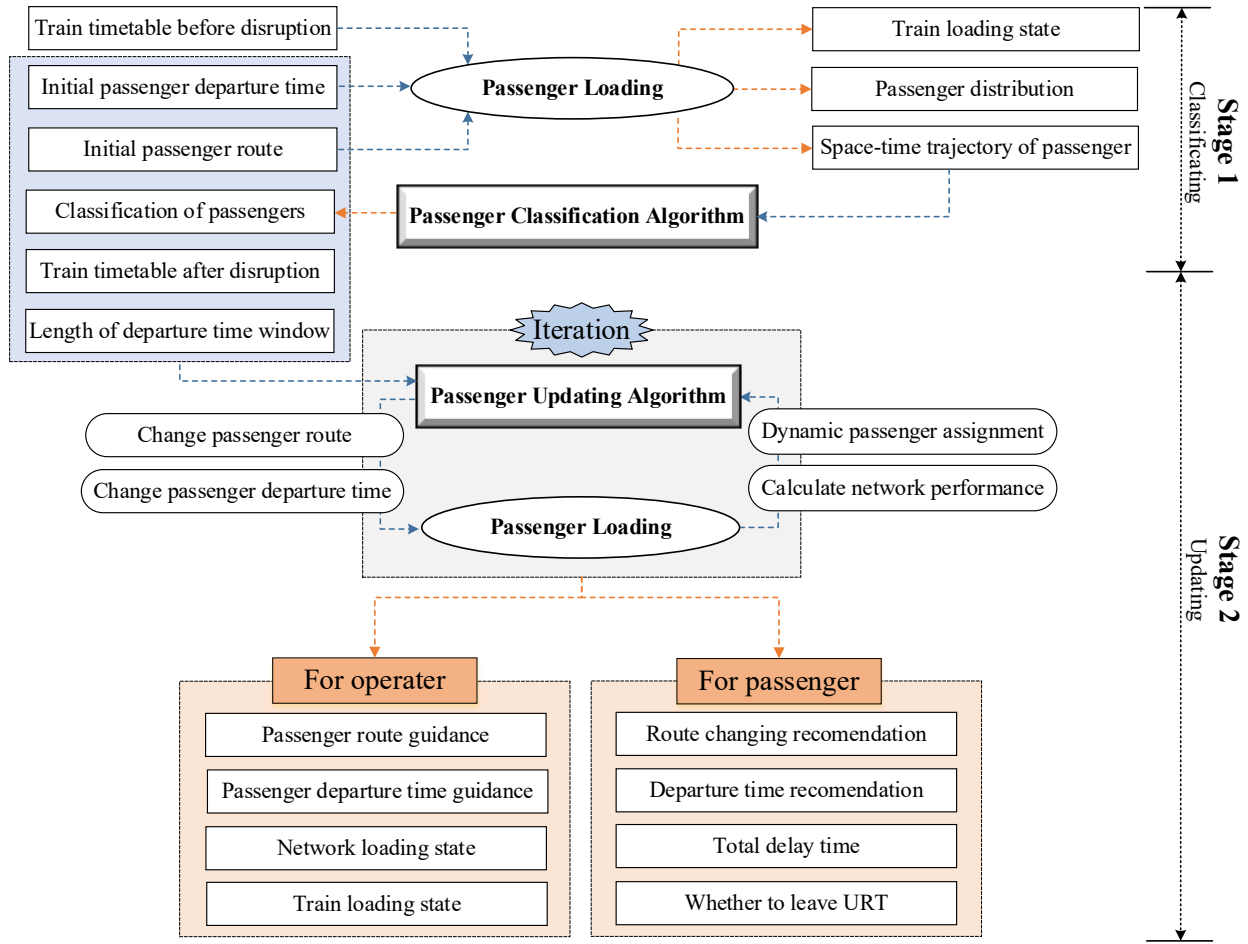

**Figure 2 Flowchart of the two-stage solution approach**

*Classification Stage*

Stage 1 primarily involves the Passenger Classification Algorithm, which occurs after the passenger loading process. Obtaining passengers' space-time trajectories is crucial for passenger classification, making the loading process necessary. Inputs for this loading process include initial travel routes and departure times. Using the train timetable before disruption, passengers are dynamically assigned to the transportation network, obtaining spatiotemporal trajectories and total travel time, which are crucial for subsequent stages. The pseudocode for this loading process is provided by **Algorithm A** in **Appendix A**.

Following this, passengers are grouped into four groups with distinct feasible path sets and departure time sets, determined by their space-time trajectories. The classification principles involve four key criteria: 1) whether passengers' travel routes include interrupted sections, 2) whether passengers arrive during the disruption, 3) whether passengers are inside the station when the disruption occurs, and 4) whether passengers depart after the disruption. These criteria align with the proposed three-feature four-group classification principle. The detailed process of the classification algorithm is illustrated in **Figure 3**.

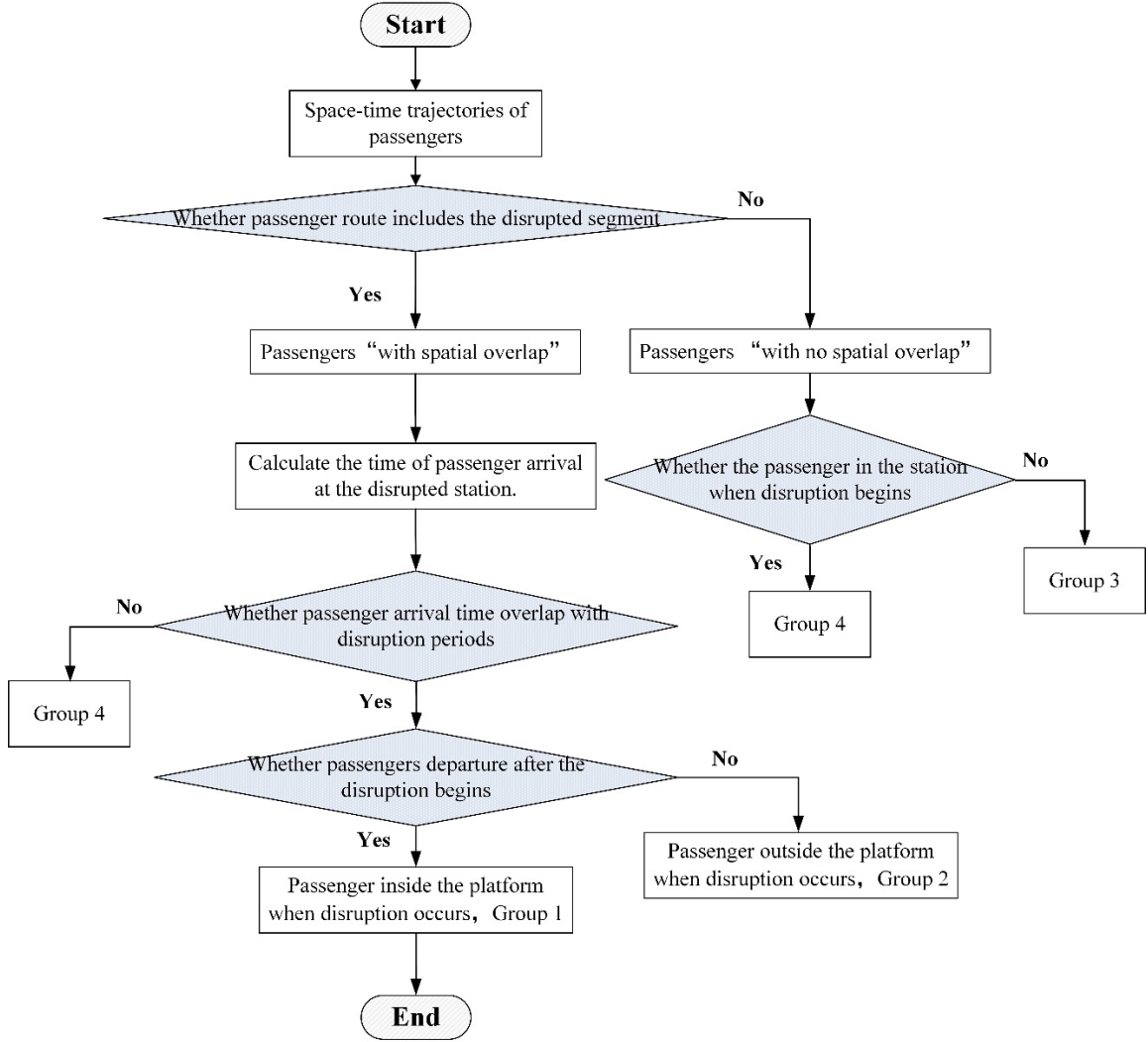

**Figure 3 The process of the passenger classification algorithm**

In practice, it is challenging to accurately estimate the number of passengers within each passenger group, in other words, the URT operator may not possess precise knowledge of the exact number of passengers stranded on a platform following a disruption. In recent years, our ongoing efforts have been focused on obtaining the states of time-dependent passenger flows in URT networks through several approaches, such as dynamic traffic assignment. Furthermore, certain studies are committed to estimating passenger flow patterns under normal operational conditions in URT systems through the use of methods such as schedule-based passenger assignments, agent-based simulations, and deep learning, among various other approaches (*21–23*). In a word, the network's states during disruptions, such as the number of passengers standing on the platform, the number of passengers transferring to another line, and in-train occupancy rates, as well as passenger locations and their spatiotemporal trajectories, can be estimated based on our preceding work (*24*).

*Updating Stage*
Stage 2 comprises the passenger updating algorithm and the passenger loading process. This stage takes the timetable after the occurrence of disruption as input. It establishes distinct updating strategies for various passenger groups derived from the passenger classification results obtained in Stage 1. The strategies involve updating passengers' travel routes and departure times while computing the passengers' total travel times to evaluate the performance of the proposed rescheduling strategies. Through an iterative process, the passenger travel routes and departure times are optimized to minimize the overall passenger travel time, thereby achieving travel rescheduling guidance for passengers under disruptions. The pseudocode for Passenger Updating Algorithm in this stage is presented in **Table 3**.

**TABLE 3 Pseudocode for passenger updating algorithm**

| Algorithm 1 Update of passengers' routes and departure times |
| --- |

**Input**:

    $\widehat{\Theta}$: Given timetable of each line after disruption

**Output**:

    $X$: Passenger assignment results

    $Z$: Choice of passengers' routes and departure times

**Parameter initialization:**

    Initialize time-dependent link travel time $LTT_{ijt} = +\infty$ for each link $(i,j)$ and time $t$

    Initialize time-dependent route travel time $PTT_{pt} = +\infty$ for each route $p$ and time $t$

    Initialize route and departure-time choice variables $z_{pt}^a = 0$ for each passenger $a$, route $p$, and time $t$

    **for** iteration step $k$ from 1 to $k^{pas}$, **do**

        get passenger assignment results $X$ by **Algorithm A** in **Appendix A**

        **for** $p \in P$, **do**

            **for** $t \in T$, **do**

                set cumulative time $\tau = t$

                **for** each link $(i,j)$ in the link sequence set $L^p$, **do**

                    **if** link $(i,j)$ is an entry/transfer link, **do**

                        set $\tau = \tau + TT_{ij}$

                        **for** running link $(j,j')$ originated from node $j$, **do**

                              **for** time $t'$ in range $(\tau, \tau + t_{wait}^{max})$, **do**

                                    **if** there is a train service on the arc $(j,j',t',t'')$ and the train can accommodate

                                    passengers $\sum_{a \in A} x_{jj't't''}^a < \theta_{jj't't''} \times cap_k$ **do**

                                        set $\tau := \tau + TT_{jj'}$

                                        **break**

                                  **else**, **do**

                                      **pass**

                  **if** link $(i,j)$ is a running/exit link, **do**

                        set $\tau := \tau + TT_{i,j}$

                set $C_{pt} = \tau$

        **for** passenger $a \in A$, **do**

            **if** $label(a) = 1$, **do**

                **for** route $p \in P^{w(a)}$, **do**

                    find route $p^*$ by $C_{p^*t} = \min\{\dots, \dots, C_{pt}, \dots, \dots | \forall p \in P^{w(a)}\}$ and set $z_{p^*t}^a = 1$

            **if** $label(a) = 2$, **do**

                **for** route $p \in P^{w(a)}$, **do**

                    **for** time $t \in \Delta(a)$, **do**

                        find route $p^*$ and time $t^*$ by $C_{p^*t^*} = \min\{\dots, \dots, C_{pt}, \dots, \dots | \forall p \in P^{w(a)}, \forall t \in \Delta(a)\}$ and

                    set $z_{p^*t^*}^a = 1$

            **if** $label(a) = 3$, **do**

| | | **for** time $t \in \Delta(a)$, **do** |
| | | | find time $t^*$ by $C_{pt^*} = \min\{..., ..., C_{pt}, ..., ... | \forall t \in \Delta(a)\}$ and set $z_{p^*t^*}^a = 1$ |
| Return | | |

## COMPUTATIONAL EXPERIMENTS

To evaluate the effectiveness of the model formulations and the two-stage solution approach, a simple network was adopted to validate the model and analyze the key parameters, a real large-scale case based on the Beijing Subway network is used to demonstrate the optimization results of the proposed model with real-world AFC data. Experiments on the models and algorithms were performed on a desktop computer with an AMD Ryzen 7 4800H @ 3.2 GHz CPU and 16.0 GB RAM.

### Passenger Guidance Information Releasing Mechanism

Passenger guidance serves as an effective demand management strategy aimed at alleviating excessive congestion in disrupted URT networks. Before conducting numerical experiments, it is crucial to introduce the information releasing mechanism for passenger guidance in URT systems. During the process of disseminating passenger guidance information, considerations such as where, when, and what type of guidance information should be released to passengers are essential focal points (*23*).

Passenger information systems, similar to variable message signs in road networks, deliver vital updates like incidents, train schedules, weather conditions, and temporary measures to rail stations (*9*). These systems incorporate electronic message boards at station entrances, providing real-time traffic updates that passengers can access via dedicated smartphone apps or directly at stations (*10*). This traffic information enables passengers to proactively plan alternate routes or modes, alleviating concerns about potential delays and improving the overall travel experience (*11*).

In this study, we assume that there is a certain period for operators to make reaction between the occurrence of a disruption and passengers receiving information of the disruption. During this time, our proposed strategy categorizes the passenger flow inside and outside the URT network into four groups. Subsequently, information of the disruption is transmitted to all passengers who have not yet commenced their journey or entered the station via smartphone apps. These notifications include recommended travel routes and departure times. Passengers already inside the station will receive recommended route information through electronic message boards. However, in real-life scenarios, passengers may not always follow the guidance they receive. For simplicity in our study setting, we assume that all passengers will adhere to the received travel guidance.

### Small Case

#### Case Description

An auxiliary simple network is utilized to facilitate the analysis. The physical network configuration, as depicted in **Figure 4 (a)**, consists of four locations denoted by nodes A, B, C, and D, interconnected by five lines: Line 1 to Line 5. The corresponding extended network, conducive to modeling, is illustrated in **Figure 4 (b)**. In **Figure 4 (b)**, each node represents an individual platform, indicating that each node is associated with only one line. Hence, given that a location in **Figure 4(a)** may be linked to multiple lines, a location in **Figure 4 (a)** may correspond to several line platforms in **Figure 4(b)**. For instance, as node A corresponds to Line 1 and Line 2, Location A is extended to Platforms 1 and 2, whereas Location D is extended to Platforms 7, 9, and 10. The transfer links connect the platforms of different lines at the same location. In this simplified network, we designate Locations A and C as the originating points for passengers, corresponding to origin nodes 11 and 12, respectively. Passengers can access Platform 1 of Line 1 and Platform 2 of Line 2 by utilizing entry links 11-1 and 11-2, respectively. Similarly, Location D is designated as the destination for passengers, who can exit the URT system and reach the destination node 13 by utilizing exit links 7-13, 9-13, and 10-13 from Platform 7 of Line 2, Platform 9 of Line 4, and Platform 10 of Line 5, respectively.

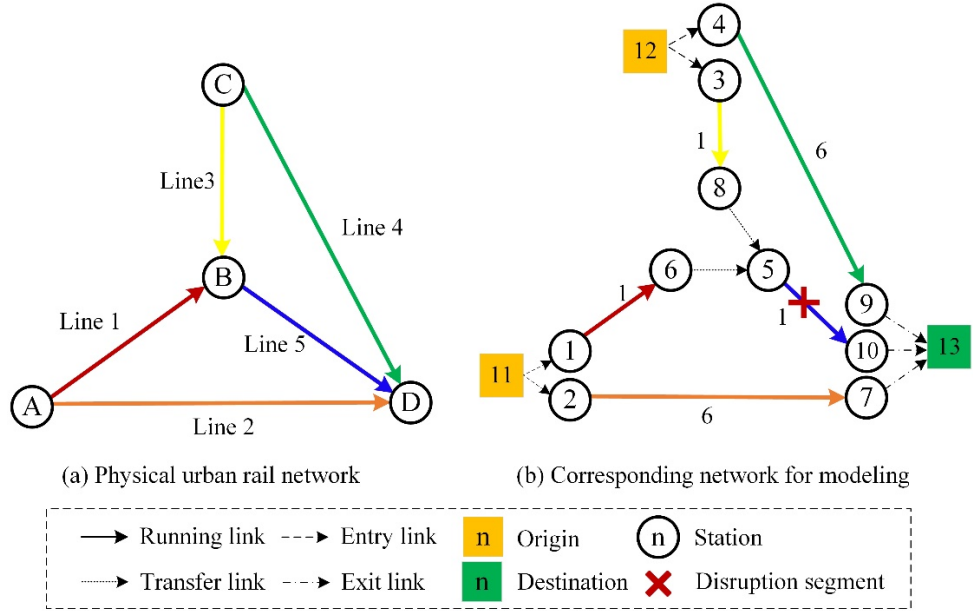

(a) Physical urban rail network      (b) Corresponding network for modeling

**Figure 4 A small simple network**

In this artificial small example, we assume that the walking time between platforms and station entrance/exits is 1 minute (including links 11-1, 11-2, 12-3, 12-4, 9-13, 10-13, 7-13). The travel time between stations on Line 1, Line 3, and Line 5 is 1 minute (i.e., links 1-6, 3-8, 5-10), while the travel time between stations on Line 2 and Line 4 is 6 minutes (i.e., links 2-7 and 4-9), and the walking time for the transfer link is 1 minute. There are two OD (origin-destination) pairs within this network: 11-13 and 12-13. Each OD pair is associated with two distinct routes. The two routes for OD pair 11-13 are a) 11-1-6-5-10-13 and b)11-2-7-13, similarly, the two routes for OD pair 12-13 are a) 12-3-8-5-10-13 and b)12-4-9-13. Based on the link travel time, we can determine that for two OD pairs, the physical travel times for the two routes are 5 minutes and 8 minutes, respectively, without considering additional factors such as transfer waiting time and congestion. Although the route that requires a transfer has a shorter physical travel time, the impact of transferring can lead people in the real world to choose a longer direct route. Taking this into consideration, we assume that the two routes for each OD pair in this example have roughly equal probabilities of being chosen by passengers.

We assumed a blockage occurred in the segment between Platforms 5 and 10 (Line 5), which meant that the operators had to stop trains at appropriate platforms (Platform 5) to wait for the blockage to be cleared. Passengers whose routes did not include the blocked location can continue their travels. However, the passengers whose travel routes involve that segment will be affected directly.

*Optimization Results*

As the input demand for this simple network, we generated 30 passengers with specific OD pairs, departure time intervals, initial paths, and the passenger groups, as shown in **Table 4**. The length of the departure time window was set to 2 min, and the studied period contained 60 intervals. The capacities of all the trains were set to 5 passengers, and the number of iterations was set to 60 in the proposed solution framework. To enhance realism in our experimental results, we assumed random arrival times for passengers and an equal 50% probability for choosing between two routes in each OD pair (i.e., Passengers randomly select their travel routes). We conducted multiple experiments and selected one instance as an example in the following description.

In a real-world application, it would take some time between when the disruption occurs and when the cause of the disruption (e.g., door jam, medical emergency, signal failure) is understood by the system's operators, allowing them to estimate a potential disruption duration and take suitable action. It would take further time for advice on changing routes or departure times to be calculated and provided to passengers, and for them to act on it. Therefore, in this small example, we assume that the time from the occurrence of disruption to passengers receiving the message about the disruption is 5 minutes. Based on this assumption, within the first 25 minutes of the study period, passengers will not be aware of any disruptions and their behavior will remain unchanged. Subsequently, the 30 passengers are categorized into 3 groups based on the classification principle proposed in this study. The number in the last column of **Table 4** corresponds to Group 1 (passengers who can only change their routes), Group 2 (passengers who can modify both their departure times and routes), and Group 4 (passengers who cannot modify either their departure times or routes). Due to the absence of passengers whose routes are without interrupted segment, Group 3 is not exit in this small case.

**TABLE 4 Information of input agents**

| Agent_id | From_node | To_node | Departure_time | Late_departure_time | Path_id | Group |
|----------|-----------|---------|----------------|---------------------|---------|-------|
| 1 | 11 | 13 | 3 | 5 | 1 | 4 |
| 2 | 11 | 13 | 6 | 8 | 2 | 4 |
| 3 | 11 | 13 | 9 | 11 | 1 | 4 |
| 4 | 11 | 13 | 12 | 14 | 2 | 4 |
| 5 | 11 | 13 | 15 | 17 | 1 | 4 |
| 6 | 11 | 13 | 17 | 19 | 1 | 1 |
| 7 | 11 | 13 | 18 | 20 | 2 | 4 |
| 8 | 11 | 13 | 21 | 23 | 1 | 1 |
| 9 | 11 | 13 | 24 | 26 | 2 | 4 |
| 10 | 11 | 13 | 27 | 29 | 1 | 2 |
| 11 | 11 | 13 | 30 | 32 | 2 | 4 |
| 12 | 11 | 13 | 33 | 35 | 1 | 2 |
| 13 | 11 | 13 | 36 | 38 | 2 | 4 |
| 14 | 11 | 13 | 39 | 41 | 1 | 4 |
| 15 | 11 | 13 | 42 | 44 | 2 | 4 |
| 16 | 12 | 13 | 3 | 5 | 5 | 4 |
| 17 | 12 | 13 | 6 | 8 | 6 | 4 |
| 18 | 12 | 13 | 9 | 11 | 5 | 4 |
| 19 | 12 | 13 | 12 | 14 | 6 | 4 |
| 20 | 12 | 13 | 15 | 17 | 5 | 4 |
| 21 | 11 | 13 | 17 | 19 | 5 | 1 |
| 22 | 12 | 13 | 18 | 20 | 6 | 4 |
| 23 | 12 | 13 | 21 | 23 | 5 | 1 |
| 24 | 12 | 13 | 24 | 26 | 6 | 4 |
| 25 | 12 | 13 | 27 | 29 | 5 | 2 |
| 26 | 12 | 13 | 30 | 32 | 6 | 4 |
| 27 | 12 | 13 | 33 | 35 | 5 | 2 |
| 28 | 12 | 13 | 36 | 38 | 6 | 4 |
| 29 | 12 | 13 | 39 | 41 | 5 | 4 |
| 30 | 12 | 13 | 42 | 44 | 6 | 4 |

We have set the time of the disruption to occur from 20 minutes after the hour to 40 minutes after the hour. **Figure 5** illustrates the changes in the timetable before and after the disruption. The red, orange, yellow, green, and blue lines in this figure respectively represent the train timetables for Line 1 to Line 5. These colors also correspond to the line colors in **Figure 4**. To make this small example more realistic, we randomly set the departure times of trains. **Figure 5** depicts one of the randomly generated scenarios, where the first trains of the five lines depart at intervals of 2, 3, 4, 5, and 2 minutes, respectively. The headways for lines 1, 2, 3, 4, and 5 are 3, 4, 5, 2, and 5 minutes, respectively.

In practice, once the disruption is resolved, the accumulated trains are typically dispatched as quickly as permitted by the train signal system, which usually results in a shorter headway than 3 minutes. This scenario also assumes no operational interventions, such as short turns, to minimize subsequent disruption to the other direction of the line. Therefore, as shown in **Figure 5(b)**, after the disruption, the headway of Line 5 is reduced from 5 minutes to 2 minutes. This increase in train frequency ensures that the total number of train services before and after the disruption remains consistent, ensuring that trains delayed at subsequent platforms due to the disruption are dispatched as promptly as possible.

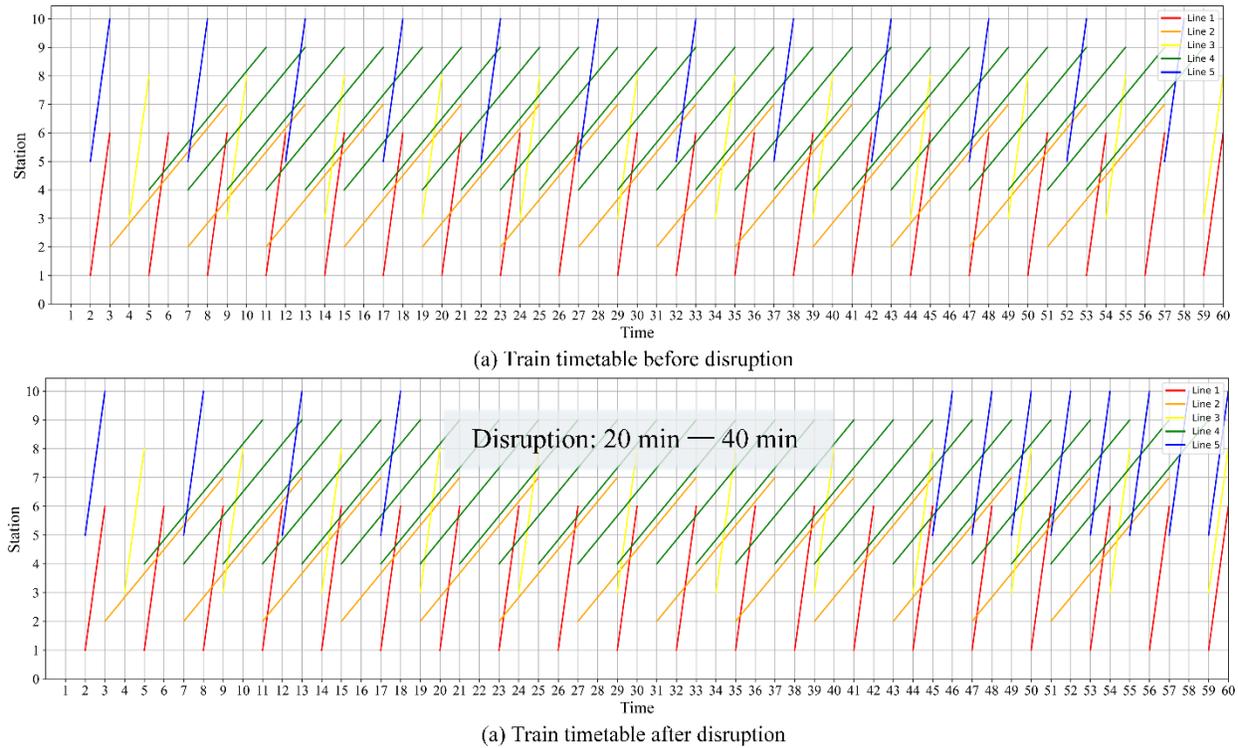

(a) Train timetable before disruption

(a) Train timetable after disruption

**Figure 5 Train timetable before and after disruption**

The proposed model and solution approach yielded results within about 1 second, devising distinct travel rescheduling strategies for the four groups of passengers based on their respective feasible decision spaces. **Figure 6** illustrates the spatiotemporal trajectories of passengers under three scenarios: (a) before disruption, (b) after disruption without passenger travel rescheduling guidance, and (c) after disruption with passenger travel rescheduling guidance. The red, green, and blue lines represent the spatiotemporal trajectories of passengers whose travel routes remain unchanged, can only change routes, and can modify both travel routes and departure times, respectively.

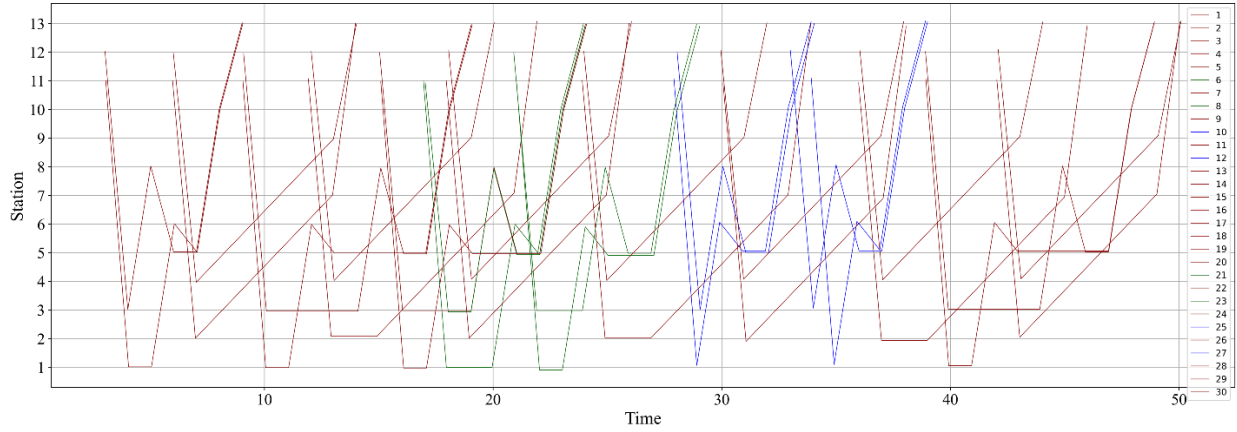

(a) Passenger trajectories without disruption

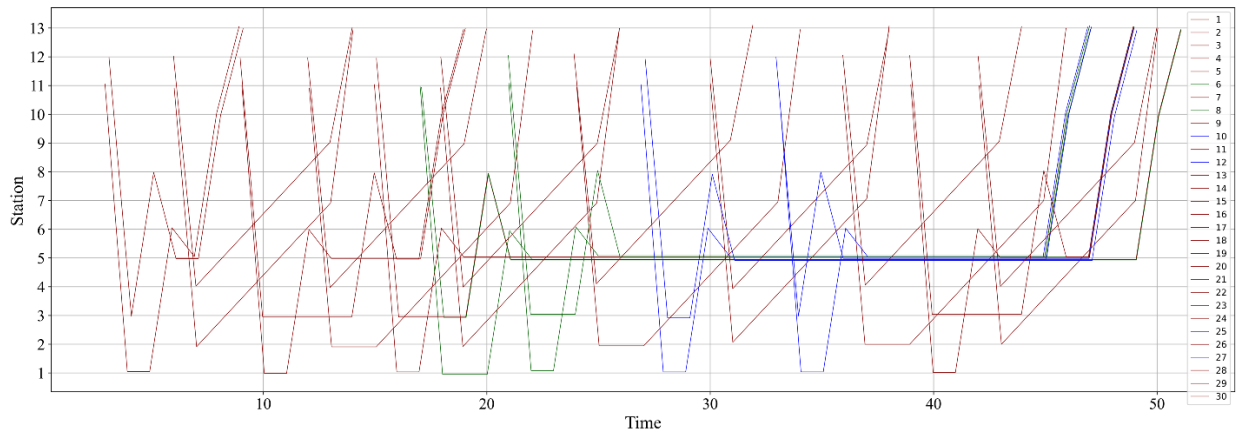

(b) Passenger trajectories under disruption without trip rescheduling guidance

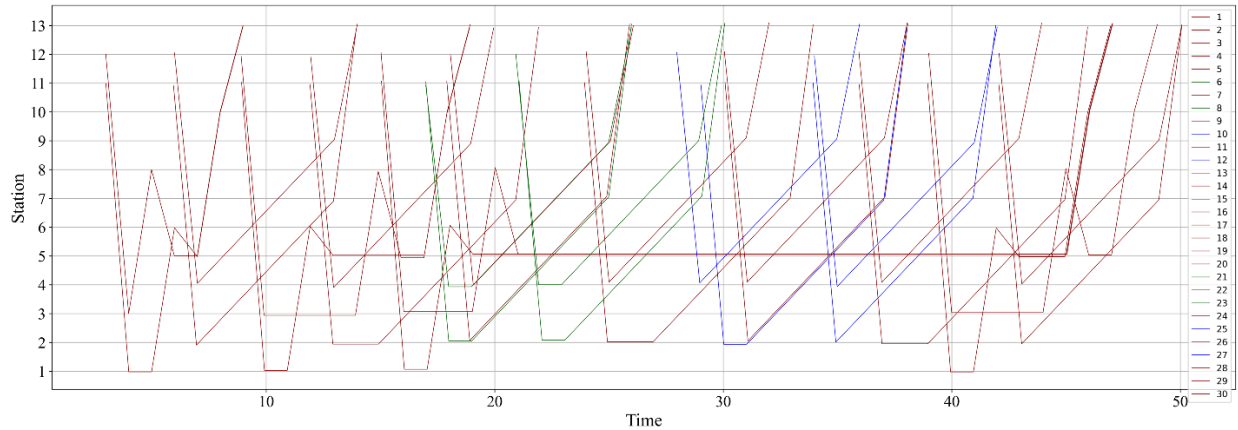

(c) Passenger trajectories under disruption with trip rescheduling guidance

**Figure 6 Passengers' space-time trajectories**

As depicted in **Figure 6(a)**, the total passenger travel time before the disruption was 241 minutes. After the disruption, passenger travel times increased to 428 minutes, as shown in **Figure 6(b)**. However, thanks to the travel rescheduling strategies proposed in this study, total passenger travel time was effectively reduced to 301 minutes, as illustrated in **Figure 6(c)**. In comparison to scenarios without travel rescheduling guidance, the proposed approach can lead to a substantial reduction of approximately 29.7% in passengers' total travel time. From **Figure 6** we can observe that, after receiving travel rescheduling guidance, the green passengers have shortened their travel time by changing their travel routes (e.g., change from 11→1→6→ $5\rightarrow13$ to $11\rightarrow2\rightarrow7\rightarrow13$). Similarly, the blue passengers have effectively mitigated the impact of unexpected disruptions by either altering their travel routes or delaying their departure times.

*Computation efficiency of the Gurobi solver and the proposed two-stage solution approach*

To assess the computational performance of our proposed two-stage solution approach, we conducted computations for five distinct scenarios, each characterized by varying passenger demand levels ranging from 30 to 3000. These scenarios were solved using both the standard optimization solver Gurobi and our proposed two-stage solution approach. **Table 5** and **Table 6** offer a comprehensive summary of the results obtained from comparing the computation time and quality performance between the two solving methods.

Within **Table 5** and **Table 6**, the columns labeled "# $x_{ijtt'}^a$," and "# constraints" present the counts of variables and constraints, respectively. The column titled "Obj" displays the average value of the objective function obtained after conducting five experiments under the same input conditions. Additionally, the column "Gap" indicates the disparity between the objective values produced by Gurobi and our proposed two-stage solution approach.

As shown in **Table 5**, we have further divided the total solving time into data loading time, model building time, and solution/CPU time. From the data results, we can observe that, our proposed two-stage solution approach significantly outperforms Gurobi in terms of computational time, achieving an impressive average CPU time reduction of 90.5%. Notably, as the scale of the problem increases, the computational time required by Gurobi escalates substantially in comparison to our two-stage solution approach.

In Case 5, wherein the passenger count reaches 3000, Gurobi fails to yield an optimal solution within an acceptable computational time threshold (e.g., 3600 seconds). In contrast, our approach demonstrates remarkable efficiency, requiring approximately 21 seconds and 60 iterations to reach a solution.

**TABLE 5 Computation time comparison of Gurobi and our two-stage solution approach**

| Case No. | # passengers | Solution approach | Data loading time | Modeling building time | Solution/CPU time | Total computation time (s) | Reduced CPU time (%) |
|---|---|---|---|---|---|---|---|
| 1 | 30 | Gurobi | 1.53 | 0.50 | 1.26 | 2.29 | 77.18% |
| | | Our approach | 0.69 | 0.1725 | 0.2875 | 1.15 | |
| 2 | 100 | Gurobi | 21.68 | 7.09 | 3.68 | 32.45 | 88.99% |
| | | Our approach | 0.972 | 0.243 | 0.405 | 1.62 | |
| 3 | 300 | Gurobi | 69.20 | 22.62 | 11.76 | 103.58 | 91.11% |
| | | Our approach | 2.508 | 0.627 | 1.045 | 4.18 | |
| 4 | 1000 | Gurobi | 359.42 | 117.46 | 61.08 | 537.96 | 96.58% |
| | | Our approach | 5.016 | 1.254 | 2.09 | 8.36 | |
| 5 | 3000 | Gurobi | 2531.52 | 827.29 | 430.19 | > 3600 | 98.77% |
| | | Our approach | 12.696 | 3.174 | 5.29 | 21.16 | |

In terms of solution quality, our proposed two-stage solution approach consistently delivers an approximate optimal value with a minimal gap when compared to results obtained from Gurobi. Across all four cases, the average gap remains at a mere 0.40%.

**TABLE 6 Computation quality comparison of Gurobi and our two-stage solution approach**

| Case No. | # passengers | # $x_{ijtt'}^a$ | # constraints | Solution approach | Obj (min) | Gap (%) |
|---|---|---|---|---|---|---|
| 1 | 30 | 31528 | 61618 | Gurobi | 301 | 0% |
| | | | | Our approach | 301 | |
| 2 | 100 | 142598 | 758694 | Gurobi | 715 | 0% |
| | | | | Our approach | 715 | |
| 3 | 300 | 434667 | 3956194 | Gurobi | 2687 | 0.67% |
| | | | | Our approach | 2705 | |
| 4 | 1000 | 1805987 | 17589874 | Gurobi | 7358 | 0.92% |
| | | | | Our approach | 7426 | |
| 5 | 3000 | 5678954 | 58975462 | Gurobi | - | - |
| | | | | Our approach | 23594 | |

*Mode Comparison and Sensitivity Analysis*

We conducted a comparative analysis of different scenarios to demonstrate the advantages of the proposed passenger guidance strategies. Specifically, we compared the variations in total travel time for passengers under four modes when disruption occurred: **Scenario 1**, where neither passenger travel routes nor departure times were adjusted; **Scenario 2**, where only departure times were adjusted; **Scenario 3**, where only passenger travel routes were adjusted; and **Scenario 4**, where both passenger travel routes and departure times were adjusted.

Compared to Scenario 1, the results indicate that adjusting only departure times can reduce total travel time by 10.7%, and adjusting only passenger travel routes can reduce total travel time by 21.3%. This demonstrates that changing passenger routes is more effective in reducing the impact of disruptions on passengers. Compared to no rescheduling, 29.7% of travel time can be saved with travel rescheduling guidance.

Simultaneously, we conducted a sensitivity analysis on various parameters, including train capacity, length of departure time window, and disruption duration. In each experimental group, all parameters except the variable were kept constant. The results are presented in **Table 7**.

As the train capacity increased (from 3 to 6) and the length of the departure time window extended (from 5 to 15), the total travel time for passengers decreased. Increasing train capacity significantly contributed to reducing passenger travel time, particularly when the train capacity was relatively low. However, when the train capacity reached 5 passengers, further capacity increases no longer exhibited substantial effects on reducing passenger travel time. While elongating the passenger departure time window could reduce total travel time, the magnitude of this effect did not show significant growth as the length of the time window expanded.

Furthermore, with the increase in the duration of the disruption, the passengers' travel time significantly increased. Moreover, the rate of travel time increment accelerated with longer disruption durations, indicating that long-lasting unrecoverable disruptions have a profound impact on passenger travel.

**TABLE 7 Results of mode comparison and sensitivity analysis for the small case**

| Mode Comparison | | |
|---|---|---|
| **Scenario** | **Settings** | **Total travel time (min)** |
| 1 | not change route nor delay the departure time | 428 |
| 2 | only delay the departure time | 382 |
| 3 | only change route | 337 |
| 4 | change route and delay the departure time | 301 |
| **Sensitivity Analysis** | | |
| **Scenario** | **Train capacity** | **Total travel time (min)** |
| 1 | 3 | 339 |

| 2 | 4 | 326 |
|---|---|---|
| 3 | 5 | 301 |
| 4 | 6 | 285 |
| **Scenario** | **Length of the departure time window** | **Total travel time (min)** |
| 1 | 5 | 298 |
| 2 | 10 | 295 |
| 3 | 15 | 292 |
| **Scenario** | **Duration of disruption** | **Total travel time (min)** |
| 1 | 10 | 239 |
| 2 | 20 | 301 |
| 3 | 30 | 386 |

**Large Case Based on Beijing URT Network**

This section discusses the experiment based on the Beijing Subway network (one of the busiest URT networks globally) that was considered for examining the proposed model and the two-stage solution approach.

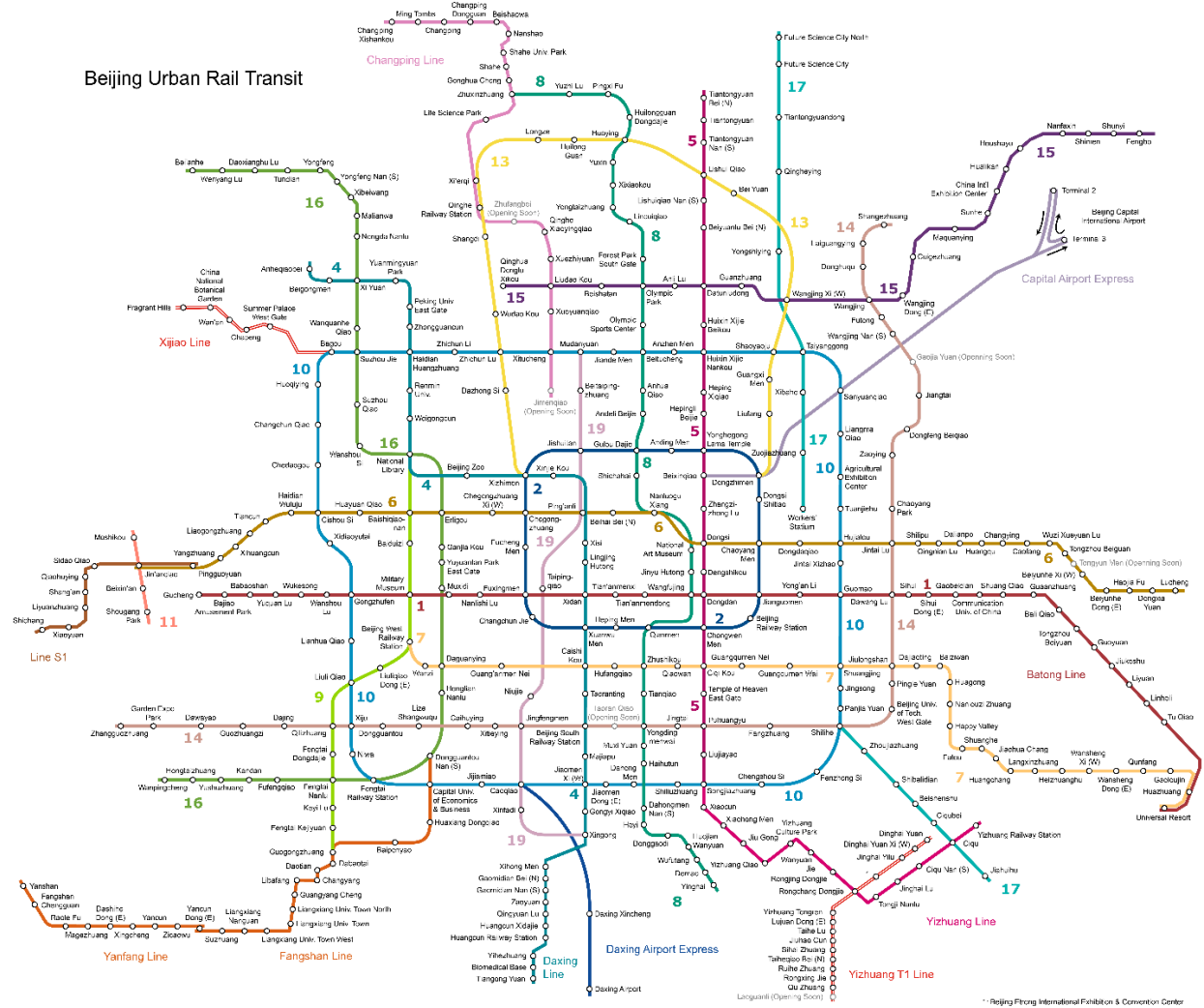

**Figure 7 Beijing Subway network (https://en.wikipedia.org/wiki/Beijing_Subway)**

As shown in **Figure 7**, the Beijing Subway network contains 27 lines and over 400 stations. We constructed an extended physical network with 1640 nodes and 3807 links to capture complex passenger waiting, riding, and transfer processes. To derive the time-dependent passenger demand for the network, 1,048,575 passenger transaction record samples were obtained from the AFC system from 7:00 to 9:00 on a typical workday; passenger origins, destinations, and departure times could be obtained from this system. The study period was from 7:00 to 9:00, with 120 1-min intervals. Additionally, we generated a candidate route pool comprising 1,000,500 potential physical routes for all the OD pairs.

As the research scope extends to such a large scale, the proposed model becomes infeasible for solutions using solvers such as **Gurobi**. In this context, the significance of the solution approach designed in this study is demonstrated. Taking the Yizhuang Line as an example, we interrupt the section between Yizhuang Bridge and Jiugong of the line for a disruption duration of thirty minutes, specifically from 8:00 to 8:30. Similarly, to minimize the impact of the disruption, the train departure headway is reduced from 6 minutes to 3 minutes after the resumption of train services. When facing unforeseen disruptions in URT systems, the effective guidance of passengers is essential to minimize travel losses for both the operators and passengers. Consequently, the rapid solving of rescheduling strategies for large-scale instances is significant. Noting that for passengers who do not require alterations to their travel routes or departure times, there is no need to design travel rescheduling schemes for them, resulting in significant time savings in solving large cases, and passengers who are not required to design a rescheduling scheme for them can be obtained by the proposed classification principle and Passenger Classification Algorithm (A2), which can assess the impact scope of disruptions quickly, determining which routes and passengers are affected. Consequently, for the Beijing subway case designed in this study, passengers are categorized by our classification algorithm into four groups: 2,217 passengers in Group 1, 2,001 passengers in Group 2, 5,385 passengers in Group 3, and 1,038,972 passengers in Group 4. Passengers in Group 4 do not require changes to their travel routes or departure times, while a total of 9,603 passengers in Groups 1, 2, and 3 do require alterations under disrupted conditions, accounting for less than 1% of the total passengers (1,048,575). During the computing process of our iterative solving framework, we can achieve a near-optimal solution with a gap of approximately 2% in approximately 7 minutes, providing a rescheduling solution for passengers necessitating route changes and corresponding departure time delays.

Similar to the small case, the following analysis was conducted to validate further the effectiveness of the proposed model and solution approach. We compared the objective function values under three scenarios: before disruption, after disruption without travel rescheduling, and after disruption with travel rescheduling. The results indicate that the total travel time for passengers before the disruption was 249,503,697. The disruption increased total travel time to 253,800,507, causing a 1.7% passenger time delay. However, when employing the model and solution approach proposed in this study to reschedule passenger travels after disruption, the total travel time was reduced to 251,613,985. Compared to the scenario without disruption, this represents only a 0.8% increase in passenger delay. Moreover, compared to the scenario without travel rescheduling strategies, it reduces delay by 50.9%, showcasing the superiority of the proposed model and solution approach in alleviating passenger delays caused by disruptions.

## CONCLUSIONS AND FUTURE DIRECTIONS

This study focuses on the issue of guiding passengers by providing travel rescheduling strategies in URT systems under unexpected disruptions to divert passengers and mitigate congestion caused by disruptions. A three-feature four-group passenger classification principle is proposed, categorizing passengers into twelve categories based on temporal, spatial, and spatio-temporal features, which is further summarized into four groups of passenger travel choices based on their decision space. A mixed integer programming model is developed to guide passengers' routes and departure times at the network level, with a refined representation of the FIFO rule under oversaturated conditions. To address large-scale problems, a two-stage solution approach is designed, encompassing the Passenger Classification stage and the Passenger Updating stage. The Passenger Classification stage incorporates the passenger loading process, which dynamically assigns passengers, and the Passenger Updating Algorithm, which is adopted to obtain the optimal passenger routes and departure times. The proposed model and solution approach are experimentally validated on both a small example and a large-scale Beijing subway network. The results demonstrate the efficacy of the proposed strategies in alleviating passenger delays caused by unexpected disruptions. Compared to scenarios without passenger travel rescheduling strategies, the proposed guidance reduces approximately 29.7% and 50.9% in passenger travel time in the small and large-scale case studies, respectively.

The proposed behavior guidance strategies have broad potential applications. For instance, they can be utilized in reservation-based travel systems, where adjusting passengers' departure times can effectively manage the supply-demand equilibrium, thus enhancing fairness and efficiency across the system (*25*). Furthermore, these strategies can alleviate congestion during peak periods by altering passengers' travel routes or adjusting their departure times (either earlier or later), thereby optimizing passenger flow within the network.

Moreover, to provide novel approaches for addressing disruptions in URT systems, future research could further integrate the supply-side perspective by optimizing the timetable on the supply side while considering line configuration schemes. This may involve setting up full-length and short-turn routing modes to enhance efficiency. Additionally, the research could explore the introduction of passenger control strategies at stations to better manage passenger flow and ensure smoother operations (*26*). Last but not least, this paper only guides passengers without modeling their choice behavior. Future research may include integrating models like the logit model to consider uncertainty in modeling passenger behavior or exploring passengers with distinct travel purposes, such as commuting and leisure, for a more robust characterization of their choice behaviors.

## ACKNOWLEDGMENTS
This research was supported by the Fundamental Research Funds for the Central Universities under Grant No.2023YJS035, the National Natural Science Foundation of China (No. 72001020 and No. 72074215), the Joint Funds of the National Natural Science Foundation of China (No. U2034208), and Beijing Natural Science Foundation (No. 8242016).

## AUTHOR CONTRIBUTIONS
Siyu Zhuo: Data processing, Software, Visualization, Writing & Reviewing; Pan Shang: Supervision, Data collection, Methodology, Software, Reviewing; Xiaoning Zhu: Supervision, Data collection, Funding acquisition; Zhengke Liu: Data processing, Writing.

## DECLARATION OF INTEREST STATEMENT
We declare that we have no financial and personal relationships with other people or organizations that can inappropriately influence our work, there is no professional or other personal interest of any nature or kind in any product, service and/or company that could be construed as influencing the position presented in, or the review of, the manuscript entitled.

## Appendix A

The loading process of passengers are illustrated as follows.

---

**Algorithm A Passenger loading algorithm**

---

**Input**:

$\widehat{\theta}$: Given timetable of each line before disruption

$\widehat{Z}$: Given passengers' route and departure-time choice solution

**Output**:

$\mathcal{E}$: Total passengers' travel time

**Parameter initialization:**

Initialize time-dependent queue list $q_{ijtt'} = []$ for each space-time entry/transfer arc $(i, j, t, t')$

Initialize time-dependent in-vehicle passenger list $o_{ijtt'} = []$ for each space-time running arc $(i, j, t, t')$

Initialize unfinished link list $L^a$ as the chosen route's link list $L^p$ for each passenger $a$ and $\hat{z}^a_{pt} = 1$

Initialize binary train service indicator $\theta_{ijtt'}$ according to $\widehat{\theta}$ for each space-time running arc $(i, j, t, t')$

Initialize flow assignment variables $x^a_{ijtt'} = 0$ for each passenger $a$ and space-time arc $(i, j, t, t')$

| | | | | | | |
|---|---|---|---|---|---|---|
| 1 | **for** $t \in T$ | | | | | |
| 2 | 1 | **for** $(i, j) \in L$ | | | | |
| 3 | 2 | 1 | **if** $(i, j)$ is an entry/transfer link | | | |
| 4 | 3 | 2 | set $t' = t + TT_{ij}$ | | | |
| 5 | 4 | 3 | 1 | **for** running link $(j, j')$ originated from node $j$ | | |
| 6 | 5 | 4 | 2 | set $t'' = t' + TT_{jj'}$ | | |
| 7 | 6 | 5 | 3 | 1 | **if** $\theta_{jj't't''} = 1$ | |
| 8 | 7 | 6 | 4 | 2 | set boarding passenger number $\pi_{board} = \min\{len(q_{ijtt'}), cap_k - len(o_{jj't't''})\}$ | |
| | | | | | **for** passenger $a$ in the top $\pi_{board}$ of the list $q_{ijtt'}$ with the condition that the first link in $L^a$ is $(j, j')$ | |
| 9 | 8 | 7 | 5 | 3 | 1 | set $x^a_{jj't't''} = 1$ |
| 10 | 9 | 8 | 6 | 4 | 2 | append $a$ to the list $o_{jj't't''}$ |
| 11 | 10 | 9 | 7 | 5 | 3 | remove $(j, j')$ from $L^a$ |
| 12 | 11 | 10 | 8 | **else** | | |
| 13 | 12 | 11 | 9 | 1 | set $q_{ijt+1t'+1} = q_{ijtt'}$ | |
| 14 | 13 | **if** $(i, j)$ is a running link | | | | |
| 15 | 14 | 1 | **for** running/ transfer/exit link $(j, j')$ originating from node $j$ | | | |
| 16 | 15 | 2 | 1 | **for** passenger $a$ in the list $o_{ijtt'}$, with the condition that the first link in $L^a$ is $(j, j')$ | | |
| 17 | 16 | 3 | 2 | 1 | set $t' = t + TT_{ij}$ | |
| 18 | 17 | 4 | 3 | 2 | set $x^a_{jj't't''} = 1$ | |
| 19 | 18 | 5 | 4 | 3 | remove $(j, j')$ from $L^a$ | |
| 20 | 19 | 6 | 5 | 4 | **if** $(j, j')$ is a running link | |
| 21 | 20 | 7 | 6 | 5 | 1 | append $a$ to the list $o_{jj't't''}$ |
| 22 | 21 | 8 | 7 | 6 | **else if** $(j, j')$ is a transfer link | |
| 23 | 22 | 9 | 8 | 7 | 1 | append $a$ to the list $q_{jj't't''}$ |
| 24 | 23 | 10 | 9 | 8 | **else** | |
| 25 | 24 | 11 | 10 | 9 | 1 | **pass** |

**Return** $\mathcal{E} = \sum_{i,j,t,t':(i,j,t,t') \in E} \sum_{a \in A} x^a_{ijtt'} c_{ijtt'}$

---